\documentclass[draft,showpacs,preprintnumbers,amsmath,amssymb]{revtex4-1}

\newcommand{\fant}[1]{\phantom{#1}}
\newcommand{\be}{\begin{equation}}
\newcommand{\ee}{\end{equation}}
\newcommand{\wdg}{\wedge}
\newcommand{\ot}{\otimes}

\begin{document}
\begin{abstract}
A conserved current for generic quadratic curvature gravitational models is defined  and it is shown that
at the linearized level it corresponds to the Deser-Tekin charges. An explicit expression for the charge for
new massive gravity in three dimensions is given. Some implications of the linearized equations are discussed.
\end{abstract}

\title{Energy definition for quadratic curvature  gravities}

\author{Ahmet Baykal}
\email{abaykal@nigde.edu.tr}
\affiliation{Department of Physics, Faculty of Science and Letters, Ni\u gde University,  51240, Ni\u gde, Turkey}
\date\today
\maketitle

\section{Introduction}
Defining a local energy density for  a given gravitating system is still a matter of discussion,
however, it is possible to define a total energy in the general theory of relativity \cite{adm-original}.
Closely following the definition of conserved charges in the general theory of relativity \cite{abbott-deser},  a new
definition of energy for the quadratic curvature (QC) models has been proposed  in \cite{deser-tekin-PRD2007}
which  will be called Deser-Tekin (DT) charges (see, also the previous definitions \cite{Deser-Tekin-2002-2003}).
In particular, the conserved charge corresponding to  the DT energy for asymptotically flat or constant
curvature geometries have desirable properties, for example, that it circumvents the  zero-energy theorem for conformal gravity
in four dimensions \cite{boulware-horowtiz-stominger}.

Sparling forms \cite{sparling} are also used to define energy in the general relativity which are  well-suited to a treatment in terms of
differential forms relative to an orthonormal coframe \cite{szabados}.
In a direction parallel to the present work, (generalized) Sparling forms for dimensionally-continued
Euler forms in arbitrary dimensions were introduced in \cite{madore}.
The Sparling forms can  also be related to the so-called Arnowit-Deser-Misner (ADM) energy \cite{adm-original} relative to a coordinate basis
and other energy-momentum pseudo-tensors such as Landau-Lifschitz energy pseudo-tensor as well \cite{straumann}.
They also play an important role in the hamiltonian formulation of General theory of relativity \cite{goldberg}
and even in the analysis of gravitational radiation in post-Newtonian approximation for theories with torsion \cite{straumann-schweizer}.
In the present work, by using the language of the exterior algebra and orthonormal coframe formulation for the  QC gravity models, a conserved quantity
will be defined  and its connection with the DT energy will be discussed.

The rest of the paper is organized as follows.
In the following section, metric QC field equations are described in a form that is useful in defining an on-shell conserved  current
and a total energy. In the subsequent section, the linearized QC equations are studied to further investigate
the relation of the energy definition presented with the DT charges.
In the last section, an explicit  expression for the conserved current in the form of an exact  differential
form for New massive gravity in three dimensions is presented.

\section{QC field equations and definition of a new conserved current}
 The notation and the conventions for the
geometrical quantities closely follow those of \cite{ugur, baykal-delice}, see also \cite{straumann}.
For the technical details related, in particular, to the QC field equations relative to an orthonormal coframe, the reader is referred to  \cite{ugur}.

A convenient starting point is to recall that the identity $*G^\alpha=-\frac{1}{2}\Omega_{\mu\nu}\wdg *\theta^{\alpha\mu\nu}$
for the Einstein form can be used to split it into the sum of two terms as
\be\label{einstein-split}
*G^\alpha
=
d *F^\alpha+*t^\alpha
\ee
where $F^\alpha$ is known  as Sparling 2-form \cite{sparling} with the expression
\be
*F^\alpha
=
-\tfrac{1}{2}\omega_{\mu\nu}\wdg *\theta^{\alpha\mu\nu}
\ee
whereas the pseudo-tensorial energy-momentum 3-form $*t^\alpha$ is explicitly given by
\be\label{LL-density}
*t^\alpha
=
\tfrac{1}{2}
(\omega_{\mu\nu}\wedge\omega^{\alpha}_{\phantom{Q}\lambda}
\wedge*\theta^{\mu\nu\lambda}
-
\omega_{\mu\lambda}\wedge\omega^{\lambda}_{\fant{a}\nu}\wedge
*\theta^{\mu\nu\alpha}
).
\ee
The standard ADM energy expression for an asymptotically flat spacetime follows from the integral of $*F^0$, namely
\be\label{adm-expression}
E_{ADM}=\frac{1}{16\pi G}\int_{\Sigma}*F^0
\ee
where $\Sigma$ is a spacelike hypersurface (see, e.g., \cite{straumann}, p. 118).
For the   well-known  example of Schwarzchild metric, by introducing the isotropic coordinates,
the formulae (\ref{adm-expression}), singling out the leading order terms in the metric asymptotically, yields the mass parameter of the metric \cite{straumann}.

As advertised above, it is possible to   define conserved current, or equivalently an exact form,  for generic QC gravity by splitting the field
equations in a way similar to (\ref{einstein-split}), cf. Eqn. (\ref{qc-full-eqn-split}) below. For the sake of clarity
for the arguments and the definitions, mainly pure QC gravity models that follow from the  general lagrangian density  of the form
\be\label{qc-lag}
\mathcal{L}_{qc}
=
aR^{2}*1+bR^{\alpha}\wdg *R_{\alpha}
\ee
will  be considered ($a,b$ are dimensionless coupling constants), confining the discussion to four and subsequently to three dimensions.
For the use of first order formalism with the constraints $\omega_{\alpha\beta}+\omega_{\beta\alpha}=0$ (metric compatibility)
and $\Theta^\alpha=0$ (zero-torsion), one starts  with the extended lagrangian
\be
\mathcal{L}_e
=
\mathcal{L}_{qc}
+
\lambda^\alpha\wdg \Theta_\alpha
\ee
where $\lambda^\alpha$ is a vector-valued 2-form imposing the dynamical constraint $\Theta^\alpha=0$.
The metric compatibility constraint can simply be incorporated into the variational procedure  by $\delta\omega_{\alpha\beta}+\delta\omega_{\beta\alpha}=0$.
Subsequently, by taking the variational derivative of $\mathcal{L}_e[\theta^\alpha, \omega^{\alpha}_{\fant{a}\beta}, \lambda^\alpha]$,
one obtains metric  equations from the coframe equations expressed in terms of the pseudo-Riemannian quantities only.
For the details of the derivation of the QC field equations relative to an orthonormal coframe refer to \cite{ugur,baykal-delice}.
It follows that the vacuum equations for the QC lagrangian density (\ref{qc-lag}) then can be written in the form
\be\label{vacuum-eqn}
*E^\alpha
\equiv
D\lambda^\alpha+*T^\alpha
=0
\ee
where the vector valued 1-form  $E^\alpha=E^\alpha_{\fant{a}\beta}\theta^\beta$ is defined by (\ref{vacuum-eqn}) and is a symmetric vector-valued
1-form (by invariance of the action under local coframe rotation).  As in the Einstein-Hilbert case, it is covariantly constant, $D*E^\alpha=0$ as a result of
diffeomorphism invariance of the QC lagrangian. Moreover, the pure QC terms are contained in $*T^\alpha$ term where
$T^\alpha=T^\alpha_{\fant{a}\beta}\theta^\beta$ is another  symmetric vector-valued 1-form.
In a generic QC model, fourth order lagrange multiplier term, $D\lambda^\alpha$,  and
second order terms,  $*T^\alpha$, are both expressible in terms of auxiliary antisymmetric-tensor-valued 2-form $X^{\alpha\beta}$. For the
QC lagrangian (\ref{qc-lag}) it is explicitly given by
\be\label{X-mu-nu-def}
X^{\alpha\beta}
=
\theta^\alpha\wdg (bR^\beta+aR\theta^\beta)-\theta^\beta\wdg (R^\alpha+aR\theta^{\alpha}).
\ee
The connection equations can be solved uniquely for lagrange multiplier and consequently, the lagrange multiplier 2-forms $\lambda^\beta$ can be
found in terms of the auxiliary 2-form as
\begin{align}
\lambda^\beta
&=
2i_\alpha\Pi^{\alpha\beta}+\tfrac{1}{2}\theta^\beta\wdg i_\mu i_\nu \Pi^{\mu\nu}
\nonumber\\
&=
2*D[bR^\beta+(2a+\tfrac{1}{2}b)R\theta^\beta]\label{general-expression-lag-mult}
\end{align}
with $\Pi^{\alpha\beta}=D*X^{\alpha\beta}$ whereas the second term in (\ref{vacuum-eqn}) has the explicit form
\be
*T^\alpha
=
\tfrac{1}{2}(\Omega^{\mu\nu}\wdg i^\alpha*X_{\mu\nu}-i^\alpha\Omega^{\mu\nu}\wdg*X_{\mu\nu}).
\ee
Note that the trace of $*T^\alpha$ is $T^{\alpha}_{\fant{a}\alpha}*1=(4-n)\mathcal{L}_{qc}$ in $n$ dimensions.
The form of the field equation (\ref{vacuum-eqn}), which has been introduced in \cite{ugur}, directly allows one to define a
conserved quantity by inspection. In order to do so, first note that the fourth order  term can simply  be written out as
\be\label{ext-covariant-lambda}
D\lambda^\alpha
=
d\lambda^\alpha+\omega^{\alpha}_{\fant{a}\beta}\wdg \lambda^\beta
\ee
which in fact separates the terms containing fourth order partial derivatives of the metric components from the lower orders relative to a coordinate basis.
This then  motivates the definition of a convenient vector-valued 1-form $\tau^\alpha=\tau^\alpha_{\fant{a}\beta}\theta^\beta$ by
\be\label{current-def}
*\tau^\alpha
\equiv
*T^\alpha
+
\omega^{\alpha}_{\fant{a}\beta}\wdg \lambda^\beta.
\ee
Thus, by making use of  the definitions (\ref{ext-covariant-lambda}) and (\ref{current-def}) in the vacuum QC  gravity equations, (\ref{vacuum-eqn})
 then leads to the compact equations as
\be\label{qc-full-eqn-split}
d\lambda^\alpha+*\tau^\alpha=0.
\ee
Consequently, the straightforward splitting renders $*\tau^\alpha$ an exact form and, by the operator identity $d^2\equiv0$,
it is also a closed form $d*\tau^\alpha=0$.
Note however  that $*\tau^\alpha$ is ,  as in the case of  the Sparling form, pseudo-tensorial since  the right hand side of (\ref{current-def})
depends explicitly on the connection 1-form. On the other hand, for  QC gravity, $\lambda^\alpha$ will always be linear
in curvature components. Therefore, the energy-momentum 3-form (\ref{current-def})
qualifies to define a conserved current for generic QC models (\ref{qc-lag}). Consequently, it is possible to define the four momentum $P^{\alpha}$ as
\be\label{charge-def}
P^{\alpha}(\Sigma)
\equiv
\int_{\partial\Sigma}\lambda^\alpha
=
-\int_{\Sigma}*\tau^\alpha
\ee
where $\Sigma$ is a space-like hypersurface  with boundary $\partial\Sigma$.
By definition, $P^{\alpha}(\Sigma)$ vanishes identically for $\partial\Sigma=\emptyset$.
The definition  of $*\tau^\alpha$ is independent of matter content in the
sense that any matter energy-momentum can simply be included in the right hand side of (\ref{current-def}) provided that it is covariantly constant.
By inspecting the QC field equations  written out relative to a coordinate basis
(see, e.g., Eqn. (3) in Ref. \cite{deser-tekin-PRD2007})), one concludes that it is not straightforward
to obtain the splitting  (\ref{qc-full-eqn-split}) relative to a coordinate basis.

\section{Linearized QC equations}
A remarkable property of the conserved current  $\lambda^\alpha$ appears to be related to the linearized form of the vacuum equations (\ref{vacuum-eqn}).
In order to establish  connection with explicit expression defining energy presented in \cite{deser-tekin-PRD2007},  let us consider the linearization of the
field equations (\ref{vacuum-eqn}) by introducing the perturbation $h_{\alpha\beta}$ to the metric $g=\eta_{\alpha\beta}\theta^\alpha\ot\theta^\beta$
with $\eta_{\alpha\beta}=diag(-+++)$. In terms of a local orthonormal basis of coframe 1-forms, they can be defined  as
\be\label{coframe-approx-def}
\theta^{\alpha}
\approx
dx^\alpha+h^\alpha_{\fant{a}\beta}dx^{\beta}.
\ee
The indices of the symmetric $h_{\alpha\beta}$ are raised and lowered by the flat Minkowski metric $\eta$ in the linear approximation.
The relevant tensorial quantities
which are linear in $h_{\alpha\beta}$ below are indicated by  a label $L$. The harmonic gauge, where the
coordinates at use satisfy $d*dx^\alpha=0$, leads to the equation $\partial^\alpha h_{\alpha\beta}-\tfrac{1}{2}\partial_\beta h^{\mu}_{\fant{a}\mu}=0$ and
the linearized curvature forms in this gauge are
\be
R^L_\alpha
=
-\Box h_{\alpha\beta} dx^\beta, \qquad R^L=-\Box h^{\alpha}_{\fant{\alpha}\alpha}
\ee
where $\Box\equiv\eta^{\alpha\beta}\partial_\alpha\partial_\beta$ for convenience \cite{gauge}.
From these expressions, it follows that the linearized Einstein 1-form is $G^L_\alpha=R^L_\alpha-\tfrac{1}{2}R^Ldx^\alpha$ or equivalently
 $*G^\alpha_L=d*F^\alpha_L$.
The contracted Bianchi identity $D*G^\alpha=0$ holds at the linearized level in the form $d*G^\alpha_L=\partial_\beta G^{\alpha\beta}_{L}*1=0$.
With these preliminary considerations, the  vacuum QC equations (\ref{vacuum-eqn}) can now easily be linearized around Minkowski background.
(It is possible to linearize (\ref{vacuum-eqn})  around a constant curvature  background which is also known to solve
(\ref{vacuum-eqn}) exactly, see, e.g. \cite{ugur}. Constant curvature backgrounds will not be considered in the discussion here)
Since $*T^\alpha$ term involves only QC contributions, the linearization in this case involves only the constraint term $D\lambda^\alpha$, and
writing out   this term explicitly one finds
\be\label{qc-eqn-explicit}
*E^\alpha
=
2D*D[bR^\alpha+(2a+\tfrac{1}{2}b)R\theta^\alpha]+*T^\alpha=0.
\ee
Therefore, after evaluating the
exterior covariant derivatives in the constraint term and writing all the  terms out explicitly, $*E^\alpha=0$  then takes the form
\begin{align}
&
2d*d[bR^\alpha+(2a+\tfrac{1}{2}b)R\theta^\alpha]
+
2d*\{\omega^{\alpha}_{\fant{a}\beta}\wdg [bR^\beta+(2a+\tfrac{1}{2}b)R\theta^\beta]\}
\nonumber\\
&
\quad+2\omega^{\alpha}_{\fant{a}\beta}\wdg *d[bR^\beta+(2a+\tfrac{1}{2}b)R\theta^\beta]
+
2\omega^{\alpha}_{\fant{a}\beta}\wdg *\{\omega^{\beta}_{\fant{a}\mu}\wdg [bR^\mu+(2a+\tfrac{1}{2}b)R\theta^\mu]\}
\nonumber\\&
\quad+
\Omega^{\mu\nu}\wdg i^\alpha*[\theta_{\mu}\wdg (bR_\nu+aR\theta_\nu)-\theta_{\nu}\wdg (bR_\mu+aR\theta_\mu)]
-
i^\alpha\mathcal{L}_{qc}[a,b]=0.\label{exterior-derivatives-evaluated}
\end{align}
DT charges were defined by expanding this equation around a constant curvature or flat background.
In particular, the DT energy is obtained by linearizing the $(00)$ component of the equation.
As will explicitly be shown below, the definition of DT charge follows from the
linearization of the first term in (\ref{exterior-derivatives-evaluated}). Only the first term  in (\ref{exterior-derivatives-evaluated})
contains the terms linear in the partial derivatives of the metric components
and it is remarkable that this particular term  is an
exact differential form in the full QC field equations.
An important observation is that the  linearization of the first term is obtained from $D\lambda^\alpha$ simply
 by replacing the covariant exterior derivatives with exterior derivatives and by  adopting the linearized curvatures  and the Hodge dual.
Futhermore, by  approximating the orthonormal   coframe 1-forms as $\theta^\alpha\approx dx^\alpha$,
the QC equations linearized in metric components then can be written as
\be\label{lin-form1}
*E^\alpha_L
=
d\lambda^\alpha_L
=
2d*d[bR_{L\beta}^\alpha dx^\beta+(2a+\tfrac{1}{2}b)R_Ldx^\alpha].
\ee
In (\ref{lin-form1}) and in all the linearized equations involving Hodge dual below, it is necessary  that the Hodge dual is also to be approximated by
that of the  flat Minkowski space, for example, $\theta^\alpha\wdg *\theta^\beta\approx dx^\alpha\wdg*dx^\beta=\eta^{\alpha\beta}*1$, etc.
Now assuming that (\ref{lin-form1}) is an expression in the flat background, then
(\ref{lin-form1}) explicitly yields
\be\label{lin-form2}
*E^\alpha_L
=
2\partial_\nu\partial_\mu [bR^{\alpha}_{L\beta}+(2a+\tfrac{1}{2}b)\delta^{\alpha}_{\beta}R_L]dx^\nu \wdg *(dx^\mu\wdg dx^{\beta}).
\ee
By evaluating the inner products of the basis 1-forms indicated on the right hand side of (\ref{lin-form2}) with the help of the identity \cite{straumann}
\be\label{contraction-id}
dx^\nu \wdg *(dx^\mu\wdg dx^{\beta})
=
-\eta^{\nu\mu}*dx^\beta+\eta^{\nu\beta}*dx^\mu
\ee
one eventually ends up with
\be\label{lin-form3}
*E^\alpha_L
=
2P^{\alpha\mu}[b G^L_{\mu\beta}+(2a+b)\eta_{\mu\beta}R^L]*dx^\beta
\ee
where, following the notation in \cite{deser-tekin-PRD2007}, the linear differential operator $P_{\alpha\beta}$ with respect to Minkowski background defined by
\be\label{flat-lap-def}
P_{\alpha\beta}
\equiv
\partial_{\alpha}\partial_{\beta}-\eta_{\alpha\beta}\Box
\ee
is introduced for convenience. The crucial  expression (\ref{flat-lap-def}) for the projection operator $P_{\alpha\beta}$
results essentially  from the contractions in the identity (\ref{contraction-id}) which in turn follows from the general expression of $\lambda^\alpha$
in (\ref{general-expression-lag-mult}).
In the above notation, $P_{\alpha\beta}$ can equally be defined by the action of (flat space) operator $d*d$
on a vector-valued 1-form $\sigma^\alpha=\sigma^\alpha_{\fant{a}\beta}dx^\beta$ by
$d*d\sigma^\alpha=*\sigma'^\alpha$ with $\sigma'^\alpha\equiv P_{\mu\beta}\sigma^{\alpha\beta}dx^\mu$.
It follows from the definition $P_{\alpha\beta}$ and linearized Bianchi identity  $\partial_\alpha G^{\alpha}_{L\beta}=0$
that $P_{\alpha\mu}G^{\mu}_{L\beta}=-\Box G^L_{\alpha\beta}$ and therefore dropping the Hodge duals on both sides to write
(\ref{lin-form3}) component form, one obtains
\be
E_{\alpha\beta}^L
=
-2[b \Box G^L_{\alpha\beta}+(2a+b)(\eta_{\alpha\beta}\Box-\partial_{\alpha}\partial_{\beta})R^L].
\ee
The expression in the square brackets is precisely the left hand side of Eqn. (5) in \cite{deser-tekin-PRD2007}. As it was shown in
\cite{deser-tekin-PRD2007}, one of the crucial properties of the  (00) component  of the expression  is that it
involves only the second order time derivatives of metric as a result of the fact that  $P_{00}=-\nabla^2$ and
consequently it  corresponds to a  Poisson type potential. Therefore the DT energy is related to the asymptotical behavior curvature
rather than the asymptotical behavior of  metric.
Under certain assumptions about spatial asymptotics, (14) is therefore expected to yield results identical to DT charge
by singling out leading terms of the  curvature. Finding a  DT charge for a given (asymptotically flat) solution of the QC equations
requires the curvature to vanish as $\sim r^{-1}$ at spatial infinity and consequently,  such solutions exclude the metrics
having the asymptotic behavior  $\sim r^{-1}$ \cite{deser-tekin-PRD2007}.
Thus, one concludes that, without resorting to linear approximation, it is possible to define a conserved current for QC gravity and  the DT energy definition
 corresponds to the linearized  Lagrange multiplier term $d\lambda^0_L$ under certain assumptions about asymptotic behavior.
Finally, note  that the linearization of the current defined in (\ref{current-def}) then corresponds to $T_{\mu\nu}$ side of Eqn. (5)
in \cite{deser-tekin-PRD2007} up to a Hodge dual.

Now a brief discussion  about the consistency of linearized equations and some properties of their solution is in order. Any solution of
vacuum Einstein field  equations is also a solution of the vacuum QC  field equations (\ref{vacuum-eqn}),
that is, $D\lambda^\alpha=0$  and $*T^\alpha=0$ separately because of $X^{\alpha\beta}=0$ identically for a metric with $R^\alpha=0$ in the above notation.
However, the exact solutions of fourth order gravity has more integration constants \cite{pechlaner-sexl}
 and even the   linearized equations (and the solutions to them) have peculiar properties \cite{stelle}.

The property that the field equations are covariantly constant, $D*E^\alpha=0$, continues to hold in the linear  approximation.
Explicitly, the linearized Bianchi identity reads
\be
\partial_\alpha E^{\alpha}_{L\beta}
=
2\partial_\alpha P^{\alpha\mu}[b G^L_{\mu\beta}+(2a+b)\eta_{\mu\beta}R^L]
\ee
where $\partial_\alpha E^{\alpha}_{L\beta}=0$ identically follows from  the operator identity  $\partial_\alpha P^{\alpha\mu}=0$
by definition (\ref{flat-lap-def}). In the presence of matter field sources,  the field equations $*E_\alpha=-2*T^m_{\alpha}$
coupled with matter energy-momentum 1-form $T^m_{\alpha}=T^{m}_{\alpha\beta}\theta^\beta$ requires that at the linearized level,
$\partial^\alpha T^m_{\alpha\beta}=0$. The consistency of the linearized equations (\ref{lin-form3}) require a further differential
constraint on matter energy-momentum $T^{m}_{\alpha\beta}$ (which is not present in the full non-linear equations).
By calculating the trace of (\ref{lin-form3}) and making
use of the definition (\ref{flat-lap-def}) one finds that linearization of $*E_\alpha=-2*T^m_{\alpha}$ implies
\be\label{second-constraint-on-matter-em}
\Box T^{m}_{\alpha\beta}+\frac{2a+b}{2(3a+b)}P_{\alpha\beta}T^m=b\Box^2 G^L_{\alpha\beta}
\ee
where $T^m$ stands for the trace of the matter energy-momentum tensor.
For the subcase $b=0$,  the constraint (\ref{second-constraint-on-matter-em})  was derived    long ago  \cite{pechlaner-sexl}
to show that linearized equations for $R^2$ gravity do not support solutions with the energy-momentum tensor
of the form $T^m_{\alpha\beta}=\delta^{0}_{\alpha}\delta^{0}_{\beta}\rho({\bf x})$ for static, extended (and bounded) matter distribution $\rho({\bf x})$.
Explicitly, for the subcase $b=0$, (00) component of  (\ref{second-constraint-on-matter-em}) implies $\rho({\bf x})=0$ \cite{pechlaner-sexl}.
For the general case $b\neq0$ however, it yields $\rho({\bf x})\sim \Delta G^L_{00}$
where $P_{00}=\eta^{ij}\partial_i\partial_j\equiv\Delta$ and the Latin indices run over flat spatial coordinates.
This is in accordance with the fact that the expression inside the square bracket in (\ref{lin-form3})
can be regarded as a Poisson potential for the linear approximation of the  theory \cite{deser-tekin-PRD2007}.

\section{A Conserved current for New massive gravity}

The QC field equations (\ref{vacuum-eqn}) have the same form in any dimensions $n\geq3$.
For $a=n/4(n-1)$ and $b=1$ in (\ref{qc-lag}), the expression for the lagrange multiplier form in
(\ref{general-expression-lag-mult}) becomes proportional to Cotton 2-form.
In particular, for $n=3$,  the lagrangian density  for New massive gravity (NMG)\cite{nmg}  reads
\be\label{nmg-lag}
\mathcal{L}_{NMG}
=
-R*1+\frac{1}{m^2}\mathcal{L}_{K}
\ee
up to an overall coupling constant which has been dropped for convenience.
The parameter $m$ has necessarily the dimension  of mass and the QC part is  $\mathcal{L}_{K}=\mathcal{L}_{qc}[a=-\tfrac{3}{8}, b=1]$.
At the linearized level (\ref{nmg-lag}) is   equivalent to Pauli-Fierz lagrangian for a massive spin-2   particle.

In accordance with the above description, it is  convenient to write  the field equations for the QC part of (\ref{nmg-lag})
in the form given in (\ref{vacuum-eqn}) as $*E^\alpha_K=0$. The vacuum equations
that  follow from (\ref{nmg-lag}) then take the form
\be\label{vacuum-nmg}
*G^\alpha+\frac{1}{2m^2}*E^\alpha_K=0.
\ee
Each of the terms in (\ref{vacuum-nmg}) can be split to define an exact pseudo-tensorial 2-forms, $*\tau^\alpha_{K}$ for the QC part and $*t^\alpha$
that is given in (\ref{LL-density}) for Einstein-Hilbert part.
Consequently,  (\ref{vacuum-nmg}) can be rewritten in a Maxwell-like form as
\be\label{nmg-vacuum-split}
d*(F^\alpha+m^{-2}C^\alpha)+*\tau^\alpha_{NMG}=0
\ee
where $*\tau^\alpha_{NMG}$  explicitly reads
\be\label{nmg-energy1}
*\tau^\alpha_{NMG}
=
*t^\alpha
+
\frac{1}{m^2}(
\omega^{\alpha}_{\fant{a}\beta}\wdg *C^\beta
+
\tfrac{1}{2}*T^\alpha_K)
\ee
by making use of the fact that
$\lambda^\alpha=2m^{-2}*C^\alpha$  with  $C^\alpha$ standing for  the Cotton 2-form.
On the right hand side of (\ref{nmg-energy1}),
$*T^\alpha_K\equiv*T^\alpha[\Omega_{\mu\nu}, X^{\mu\nu}_{K}]$ with
$X^{\mu\nu}_{K}\equiv X^{\mu\nu}[a=-\tfrac{3}{8}, b=1]$.
Consistently, at the linearized level,   Eqn. (\ref{nmg-vacuum-split}) reads
\be\label{l-nmg-eqn}
d*(F_L^\alpha+m^{-2}C^\alpha_L)=0
\ee
(also assuming that at the linearized level $*$  stands for Minkowski Hodge dual)
and (\ref{l-nmg-eqn}) yields the field equation for a propagating massive spin-2 particle \cite{nmg}.

It is worth  to emphasize that  in (\ref{nmg-vacuum-split})
the $*F^0$ component of the Sparling form (the  asymptotically dominant term) is related to ADM energy
whereas $\lambda^0=m^{-2}*C^0$ component is related to DT energy definition for generic higher curvature gravity.
The new energy definition in Section II allows one to combine these  energy definitions for NMG in a natural way.

As mentioned in the introduction, one of the motivations for the construction of the DT charges is to solve the zero
energy problem of Weyl gravity with flat background. Weyl gravity follows from the lagrangian $\mathcal{L}_{qc}[a=-\frac{1}{3},b=1]$ in four dimensions
and  in this case, $\lambda^\alpha=2*C^\alpha$ as in NMG \cite{ugur}. Consequently, in four dimensions  $P^\alpha$ in (\ref{charge-def})
does not vanish identically in accordance with  the DT result.
On the other hand, for $n\geq 4$ dimensions, $\lambda^\alpha=0$ identically for Gauss-Bonnet lagrangian density
$\Omega_{\alpha\beta}\wdg\Omega_{\mu\nu}\wdg *\theta^{\alpha\beta\mu\nu}$
 and the corresponding field equations are second order in metric components.
 Consequently, in this case  $P^0$ in (\ref{charge-def})  yields zero identically for the energy
 as the DT definition of energy does as well.

The author would like to thank \"Ozg\"ur Delice for helpful comments and critical reading of the manuscript.


\begin{thebibliography}{99}

\bibitem{adm-original}
R Arnowitt, S Deser and  C W Misner,
Dynamical Structure and Definition of Energy in General Relativity,
Phys. Rev. {\bf116}, (1959) pp. 1322-1330

\bibitem{abbott-deser}L F Abbott and S Deser,
Stability Of Gravity With A Cosmological Constant,
Nuclear Physics B {\bf195} (1982) pp. 76-96

\bibitem{deser-tekin-PRD2007}S Deser and Bayram Tekin,
New energy definiton for higher curvature gravities, Phys. Rev. D {\bf75} (2007) 084032

\bibitem{Deser-Tekin-2002-2003} S Deser and Bayram Tekin,
Gravitational energy in Quadratic-Curvature Gravities, Phys. Rev. Lett. {\bf89} (2002) 101101;
S Deser and Bayram Tekin,
Energy in generic higher  Curvature Gravity theories,
Phys. Rev. D {\bf67} (2003) 084009


\bibitem{boulware-horowtiz-stominger}D G Boulware, G T Horowitz, and A Strominger,
Zero-Energy Theorem for Scale-Invariant Gravity,
Phys. Rev. Lett. {\bf50}, (1983) pp. 1726-1729

\bibitem{sparling}G A J Sparling,
Twistors, spinors and the Einstein vacuum equations, University of Pittsburg, \emph{Preprint}, (1984);
W Thirring, Classical Mathematical Physics:  Dynamical Systems and Field Theories,
3rd Edition (Springer, New York, 2003)


\bibitem{szabados} L B Szabados,
On canonical pseudotensors, Sparling's form and Noether currents,
Class. Quantum Grav. {\bf9} (1992) 015014;
J\"org Frauendiener, Geometric description of energy-momentum preudotensors,
Class. Quantum. Grav. {\bf 6} (1989) pp. L237-L241

\bibitem{madore} Michel Dubois-Violette and John Madore,
Conservation laws and integrability conditions for gravitaional and Yang-Mills Field equations,
Commun. Math. Phys., {\bf108} (1987) pp. 213-223

\bibitem{straumann}N Straumann,
General Relativity: With Applications to Astrophysics (Theoretical and Mathematical Physics),
Springer, (2010)

\bibitem{goldberg}J N Goldberg,
Triad approach to the Hamiltonian of general relativity, Phys. Rev. D,  {\bf37} (1988) pp. 2116-2120;
R P Wallner,
New variables in gravity theories, Phys. Rev. D {\bf42},  (1990) pp. 441-448

\bibitem{straumann-schweizer}
M Schweizer, N Straumann and A Wipf,
Post-Newtonian Generation of Gravitational Waves in a theory of gravity with torsion,
Gen. Relativ. Gravit.,   {\bf12}  (1980) pp. 951-961

\bibitem{ugur} A Baykal,
Quadratic curvature gravity with second order trace and massive gravity models in three dimensions,
Gen. Relativ. Gravit.,   {\bf44}  (2012) pp. 1993-2017

\bibitem{baykal-delice}A Baykal and \"O Delice,
A unified approach to variational derivatives of modified gravitational actions,
Class. Quantum Grav. {\bf28} (2011) 015014

\bibitem{gauge}Transverse-traceless (TT) gauge is adopted in \cite{deser-tekin-PRD2007}. TT gauge condition implies harmonic
 gauge condition but the converse not always holds.

\bibitem{pechlaner-sexl}E Pechlaner and R Sexl,
On quadratic lagrangians in general relativity, Commun. Math. Phys. {\bf2} (1966) pp. 165-175

\bibitem{stelle}K S Stelle,
Classical gravity with higher derivatives, Gen. Relativ. Gravit. {\bf9}, (1978) pp. 353-371.

\bibitem{nmg} Eric A Bergshoeff, Olaf Hohm and Paul K Townsend,
Massive Gravity in Three Dimensions, Phys. Rev. Lett. {\bf102} (2009) 201301




\end{thebibliography}
\end{document}